\documentclass{appolb}
\usepackage{graphicx}
\usepackage[numbers,sort&compress]{natbib}
\usepackage{amsmath}
\usepackage{hyperref}
\usepackage{enumerate}
\usepackage{tikz}
\usepackage{wrapfig}
\usepackage{mathtools}
\usepackage{amssymb}
\usepackage{amsfonts}
\usepackage{physics}
\usepackage{xspace}
\usepackage{bm}
\usepackage{mathrsfs}
\newcommand{\cm}{\ensuremath{\mbox{cm}}\xspace}
\newcommand{\GeVcc}{\ensuremath{\mbox{Ge\kern-0.1em V}\!/\!c^2}\xspace}

\newcommand{\GeVc}{\ensuremath{\mbox{Ge\kern-0.1em V}\!/\!c}\xspace}
\newcommand{\NASixtyOne}{NA61\slash SHINE\ }
\newcommand{\pT}{\ensuremath{p_\text{T}}\xspace}
\newcommand{\xF}{\ensuremath{x_\text{F}}\xspace}

\date{\today}

\begin{document}
\title{Lambda transverse polarization in \NASixtyOne at the CERN SPS: feasibility studies}
\author{Yehor Bondar
\address{Institute of Physics, Jan Kochanowski University, 25-406 Kielce, Poland}
\\[3mm]
Wojciech Florkowski
\address{Institute of Theoretical Physics, Jagiellonian University, 30-348 Krak\'ow, Poland}
}
\maketitle
\begin{abstract}
The spin polarization of $\Lambda$ hyperons produced in inclusive reactions with unpolarized protons on unpolarized targets has been studied for almost 40~years. The \NASixtyOne experiment at the CERN SPS has a great potential to study transverse polarization in p-p and p-A collisions. In this work, we discuss the impact of magnetic field and limited detector acceptance on the possible polarization measurement by this experiment. Our analysis shows that the magnetic field impact on the $\Lambda$ polarization due to precession is one order of magnitude smaller than the detector acceptance-based polarization bias, and the identification of an experimental signal similar to that observed before by other experiments is possible. 
\end{abstract}

\section{Introduction}

The spin polarization of $\Lambda$ hyperons produced in inclusive reactions with unpolarized protons on unpolarized targets has been already studied over a wide range of reaction energies and various production angles of the $\Lambda$ hyperons \cite{erhan1979lambda0,PhysRevLett.36.1113,PhysRevD.91.032004,ABT2006415,Fantietal.1999,RAMBERG1994403,PhysRevD.40.3557,PANAGIOTOU1990}. Several theoretical models have been proposed to describe experimental data~\cite{ANDERSSON1979417, SZWED1981403, PhysRevD.24.2419}, however, the mechanism of polarization is still not well understood. In this work, we analyze the impact of magnetic field and limited detector acceptance on the possible polarization measurement by the \NASixtyOne experiment~\cite{Abgrall:2014fa} at the CERN SPS.

\begin{figure*}[t]
\begin{center}
\includegraphics[width=0.75\textwidth]{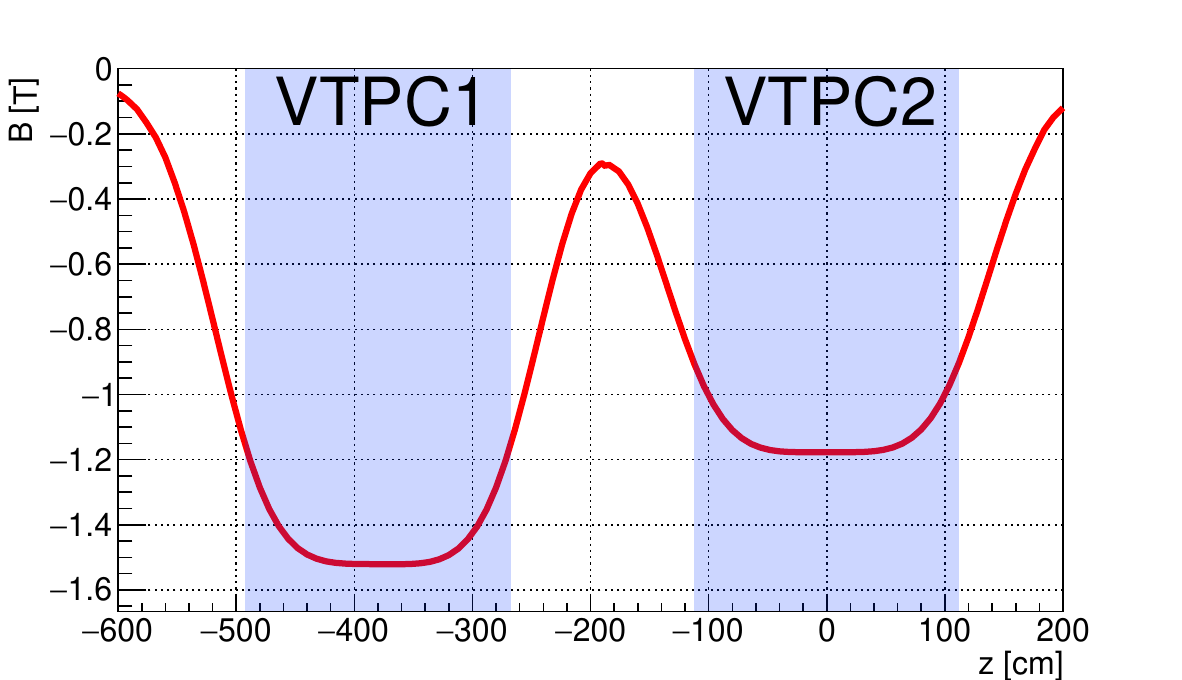}
\end{center}
\caption{
Magnetic field along the beamline ($z$) direction in the \NASixtyOne reference frame. Here $x=y=0$ and the field has only the (negative) $y$-component.
}
\label{magnetic-field}
\end{figure*}

\NASixtyOne uses two super-conducting magnets. The standard current setting for data taking at the beam momentum of 158 \GeVc corresponds to the full field of 1.5 T in the first magnet and the reduced field of 1.1 T in the second magnet, with a maximum total bending power up to 9~Tm. Magnetic field magnitude along beamline is shown in Fig.~\ref{magnetic-field}. 

The NA61/SHINE Collaboration already published the results on $\Lambda$ production in p+p interactions at the beam momentum of $158~\GeVc$~\cite{Aduszkiewicz2016}.
About 50~million proton-proton collision events were recorded in the years 2009--2011, providing an opportunity to measure $\Lambda$ transverse polarization at the previously unstudied center-of-mass energy $\sqrt{s}=$~17.3~GeV.
The \NASixtyOne spectrometer has coverage of $\Lambda$ in the range of transverse momentum $\pT\in(0,1.6)~\GeVc$ and the Feynman variable $\abs{\xF} \lesssim 0.4$~\cite{Aduszkiewicz2016},
which is the region with maximal magnitude of the expected polarization.

To study the properties of $\Lambda$ hyperons, in this work $1.2 \cdot 10^8$ events of p+p inelastic collisions with 158 \GeVc beam momentum were generated by the Epos1.99 model (version CRMC 1.4) \cite{Werner:2005jf,Pierog:2009zt}, and subsequently passed through a detector simulation employing the Geant3 package \cite{Geant3}. Simulated p+p interactions are placed uniformly at $z\in [-590,-570] \text{ cm}$ and 3 cm in diameter around $(x,y)=(0,0)$ to reproduce a $20$ cm long (2.8 \% of nuclear interaction length) liquid hydrogen target (LHT) position in \NASixtyOne detector setup. As a result,  $8.7 \cdot 10^6$ $\Lambda$s with $p \pi^-$ decay channel were analyzed. 

Hereinafter the LAB frame denotes the NA61/SHINE frame, while the rest frame refers to the $\Lambda$ rest frame. Natural units $\hbar = c = 1$ are commonly used, except for several places where $c$ is explicitly displayed.

\section{Transverse polarization -- effects of cuts}

The production plane coordinate system is defined by three mutually orthogonal unit vectors: $\hat{n}_z$ directed along $\Lambda$ momentum, $\hat{n}_x$ is transverse to beam and $\Lambda$ momenta,  $\hat{n}_y$ forms right-handed system,
\begin{equation}
    \hat{n}_x = \frac{\vec{p}_{\text{beam}}\times \vec{p}_\Lambda} {\lvert\vec{p}_{\text{beam}}\times \vec{p}_\Lambda\rvert}, \quad
    \hat{n}_z = \frac{\vec{p}_\Lambda} {\lvert\vec{p}_\Lambda\rvert}, \quad
    \hat{n}_y = \hat{n}_z \times \hat{n}_x,
          \label{eq:prodplane}
\end{equation}
where $\vec{p}_\Lambda$ is the $\Lambda$ momentum in LAB, and $\vec{p}_\text{beam}$ is the incident proton beam  momentum. In order to obtain transverse polarization, the following steps are required:
\begin{itemize}
    \item rotation from LAB to the production plane coordinate system (\ref{eq:prodplane}),
    \item boost along $\hat{n}_z$ to the $\Lambda$ rest frame, with the Lorentz transformation parameters $\beta = \abs{\vec{p}_\Lambda} / E_\Lambda$ and $ \gamma = (1-\beta^2)^{-1/2} = E_\Lambda / m_\Lambda$, where $E_\Lambda$ is the $\Lambda$ total energy in LAB and $m_\Lambda = 1.115~\GeVcc$ is the $\Lambda$ mass \cite{PDG2024},
    \item calculation of the cosines of the angles between the vector $\vec{p}_p^{\,*}$ and the directions of the three axes,
    
    \begin{equation}
        \cos\theta_i = \frac{p^{\,*}_{p\,i}}{\abs{\vec{p}_p^{\,*}}},\,\,\,i=x,y,z,
    \end{equation}
    
    \item fitting the $\cos\theta_i$ distributions and extraction of the polarization vectors $P_i$,  using a theoretically predicted distribution of proton momentum direction $f(\cos\theta_i)$   
            \begin{equation}
                \label{lambda-distribution}
            f(\cos\theta_i) = \dfrac{1+\alpha P_i \cos\theta_i}{2},
            \end{equation}
            where $\alpha = 0.732\pm0.014$ is the asymmetry parameter of the parity-violating weak decay of
the $\Lambda$ hyperon~\cite{PDG2024}.
\end{itemize}

For simulations with unpolarized protons and without any microscopic mechanism for the spin $\Lambda$ polarization (the case of Epos), one expects that $f(\cos\theta_i) = 0.5$. However, due to a finite detector coverage in space not all $\Lambda$ decays can be recorded and this introduces a bias. To include the \NASixtyOne geometry in our analysis, the following selection $\Lambda$ track cuts were imposed:

\begin{itemize}
        \item number of hits in VTPC's $\ge15$ for both proton and pion tracks\footnote{During the reconstruction procedure, some minimum amount of the tracks' hits are required for good momentum resolution.},
        \item for the difference between the $z$ coordinate of $\Lambda$ vertex (decay point) and the primary vertex, $\Delta z = z_\Lambda - z_\text{PV}$, a rapidity\footnote{\small The rapidity is defined in the standard way as $y= \text{tanh}^{-1}(p_L / E)$, where $p_L$ and $E$ are longitudinal momentum and energy of the  $\Lambda$ in the center-of-mass frame.} dependent cut was applied: $\Delta z > 10$ cm for $y<0.25$, $\Delta z > 15$ cm for $0.25\leq y \leq 0.75$, $\Delta z > 40$ cm for $0.75 \leq y \leq 1.25$, and $\Delta z > 60$ cm for $y>1.25$.
\end{itemize}
We note that these cuts are analogous to the cuts used in Ref.~\cite{Aduszkiewicz2016}.

\begin{figure*}[t]
\begin{center}
\includegraphics[width=.5\textwidth]{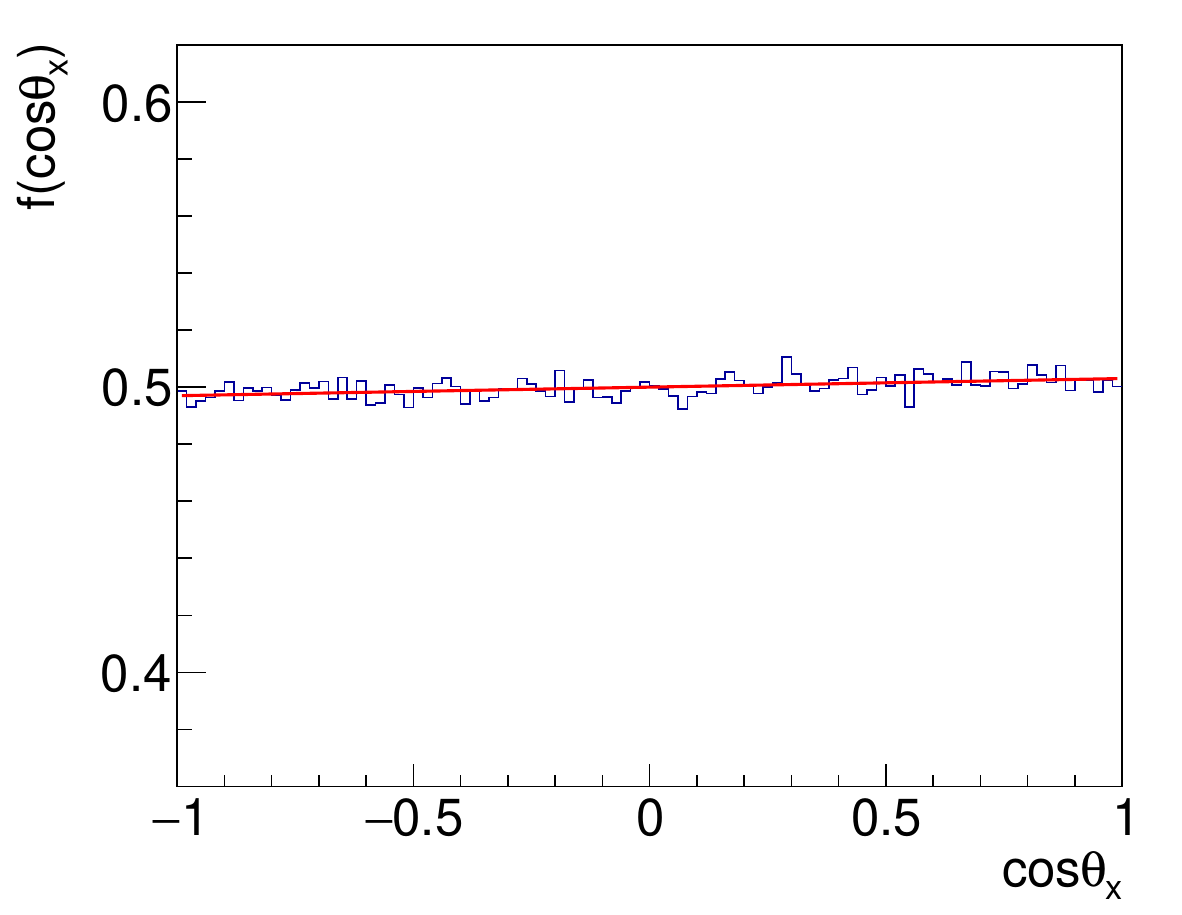} \\
\includegraphics[width=.5\textwidth]{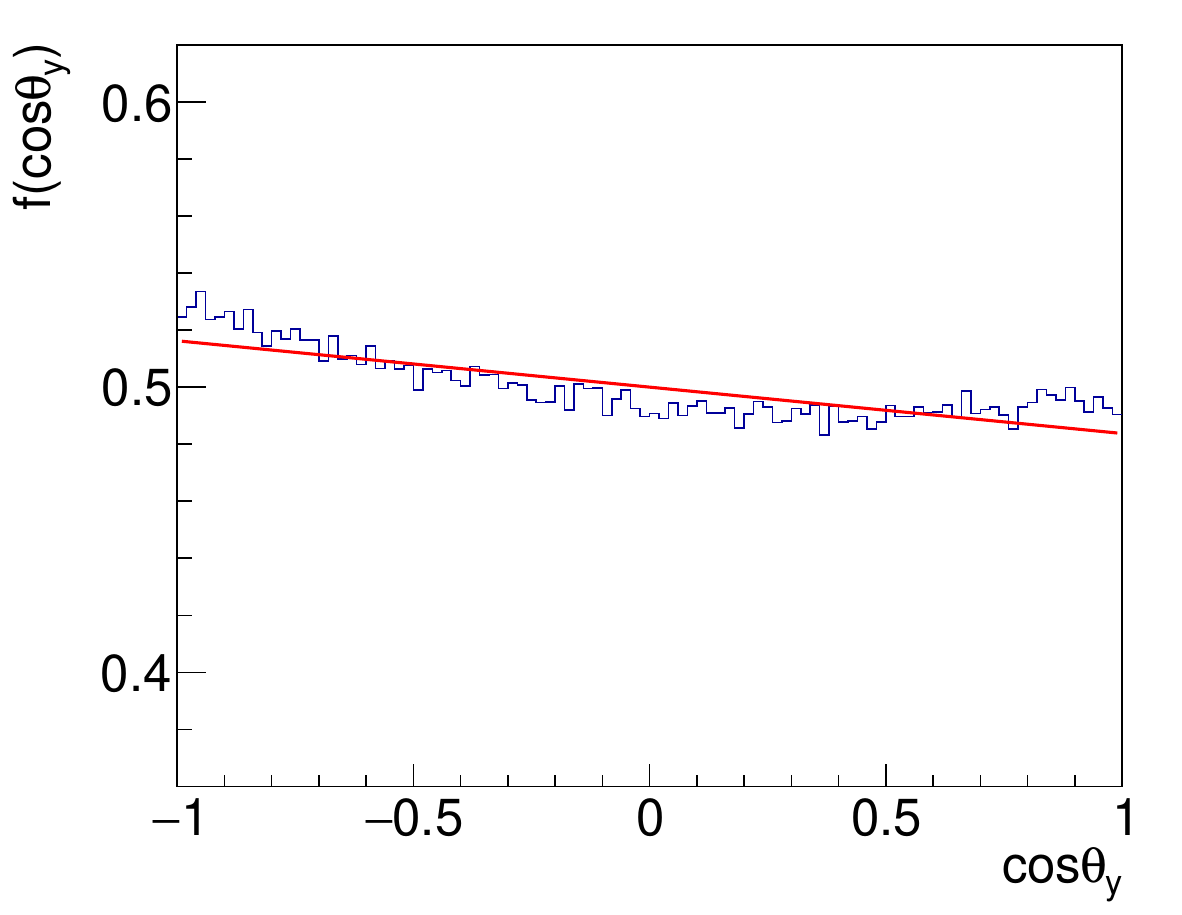} \\
\includegraphics[width=.5\textwidth]{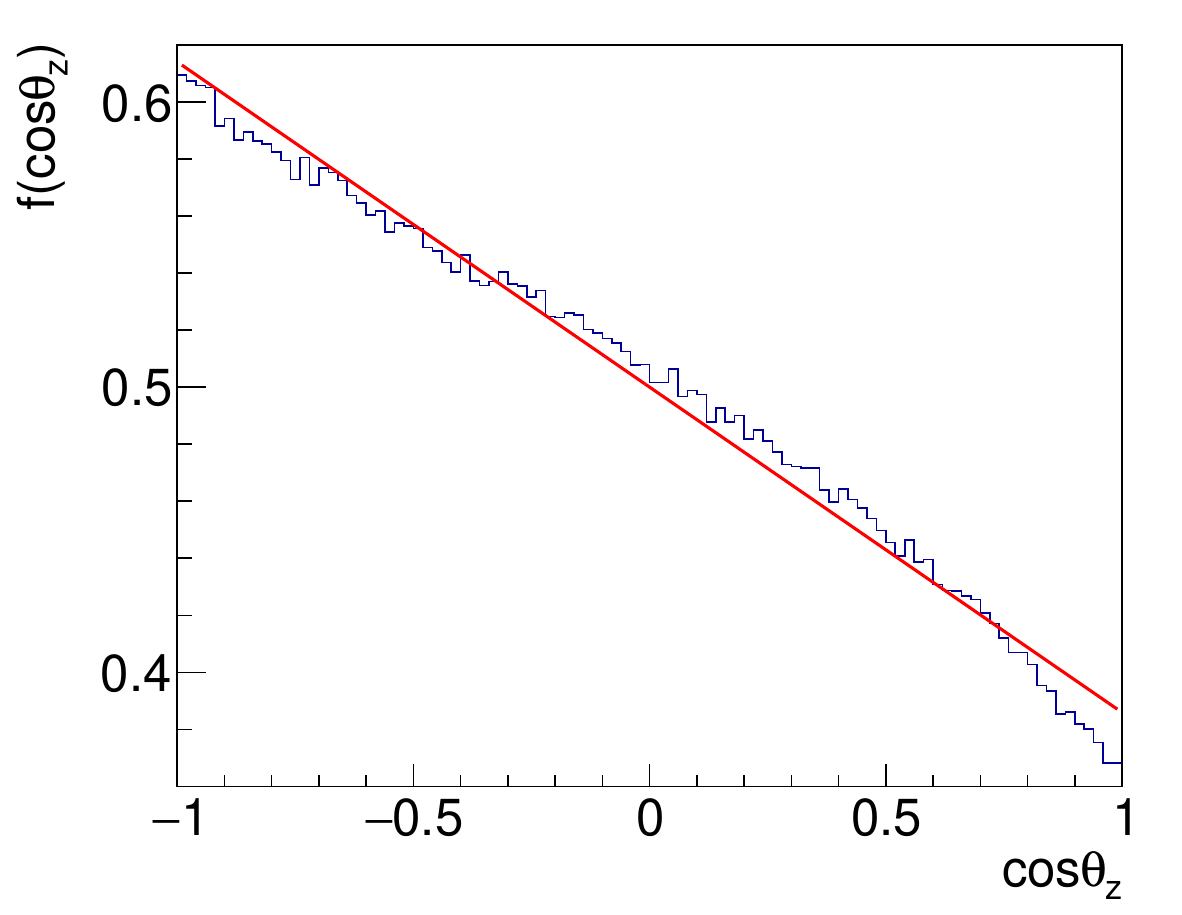}
\end{center}
\caption{
Angular distributions after application of the cuts specified in the text. Fitting to the function \eqref{lambda-distribution} gives $P_x = -0.01$, $P_y = -0.04$, and $P_z = -0.31$.
}
\label{hist-cos-cut3}
\end{figure*} 

The angular proton distributions resulting from the application of the cuts defined above to initially uniform distributions are shown in Fig.~\ref{hist-cos-cut3}. The largest deviations from 0.5 are seen for $P_z$.

\section{Lambda precession in a magnetic field (classical treatment)}

As the $\Lambda$ hyperon has a nonzero magnetic moment, it does interact with the magnetic field inside the Time Projection Chambers. Consequently, one has to investigate and quantitatively estimate whether a magnetic field can influence the polarization measurements (in addition to the cuts studied earlier). 

The classical covariant equation describing the spin motion, known as the Bargmann–Michel–Telegdi (BMT) equation, has the following form~\cite{Jackson}
\begin{align}
\label{jackson-spin-eq}
\frac{dS^\alpha}{d\tau} = \frac{ge}{2m}\left[F^{\alpha\beta}S_\beta + u^\alpha \left( S_\lambda F^{\lambda \mu} u_\mu \right)\right] - u^\alpha\left(S_\lambda \frac{du^\lambda}{d\tau}\right),   
\end{align}
where $S^\alpha$ is the Pauli-Luba\'nski spin 4-vector, $u^\alpha$ is the 4-velocity of a particle, and $\tau$ is the proper time. Here $\frac{ge}{2m} = \mu_\Lambda \mu_N$, where $\mu_N$ is the nuclear magneton and $\mu_\Lambda = -0.613$ is the magnetic moment of $\Lambda$ in $\mu_N$ units~\cite{PDG2024}. 
As gradient force terms like $\nabla(\vec{m}\cdot \vec{B})$ are negligible (see Appendix) and there are no other forces due to the fact that $\Lambda$ is a neutral particle, the last term in Eq.~\eqref{jackson-spin-eq} can be ignored. Hence, in the $\Lambda$ rest frame the spin evolution is described by the formula
\begin{equation}
\label{nonrl-spin}
\frac{d\vec{S}}{d\tau} = \mu_\Lambda \mu_N \left[ \vec{S} \times \vec{B^\prime}\right],
\end{equation}
where $\vec{B^\prime}$ is the magnetic field in rest frame in terms of the \NASixtyOne magnetic field $\vec{B}$ (electric field in LAB has been neglected \footnote{Electric field in TPC is $E=195$ V/cm, which gives $E/(cB) \approx 4\cdot 10^{-5}$.})
\begin{equation}
    \vec{B^\prime} = \gamma \vec{B} - (\gamma-1)(\vec{B}\cdot \vec{p_\Lambda}) \vec{p_\Lambda}/ \abs{\vec{p_\Lambda}}^2 . 
\end{equation}

To compute the angle of precession, we numerically integrate Eq.~\eqref{nonrl-spin}. The spatial dependence of the magnetic field strength $\vec{B}(\vec{r})$ is taken from the \NASixtyOne software package. For further convenience, we consider the change of variables to $dz = \frac{p_z}{m} c d\tau$ and rewrite Eq.~\eqref{nonrl-spin} as
\begin{align}
\label{spin-precession-dz}
    \frac{d \vec{S}}{dz}= \frac{\mu_\Lambda \mu_N}{ (p_z / m)}
    \left[ \vec{S}\times \vec{B^\prime}(\vec{r})\right].
\end{align}
Finally, the integration over the $z$ coordinate with step $\Delta z = 1\text{ cm}$ is made. 

\begin{figure*}[t]
\begin{center}
\includegraphics[width=0.6\textwidth]{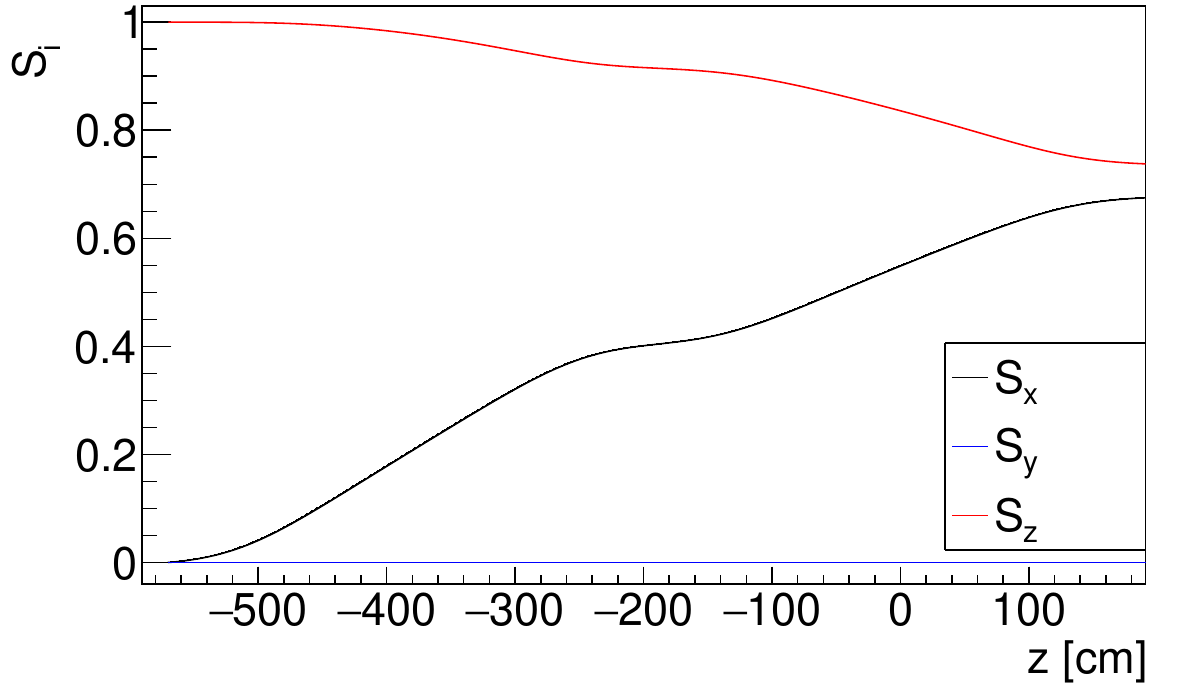}
\end{center}
\caption{
An extreme example showing the evolution of the spin vector components of the $\Lambda$ hyperon in the \NASixtyOne spectrometer magnetic field for the momentum $\vec{p}_{\Lambda}=(0,0,40)$~\GeVc.
}
\label{xyzspin-sample}
\end{figure*}

An example of the space evolution of the three spin vector components of $\Lambda$ in the \NASixtyOne spectrometer magnetic field for the momentum $\vec{p}_{\Lambda}=(0,0,40)$~\GeVc is shown in Fig.~\ref{xyzspin-sample}. In this case the distance travelled by $\Lambda$ is quite substantial due to the time dilation effect caused by the large $\Lambda$ momentum.

\begin{figure*}[htbp]
\begin{center}
\includegraphics[width=.6\textwidth]{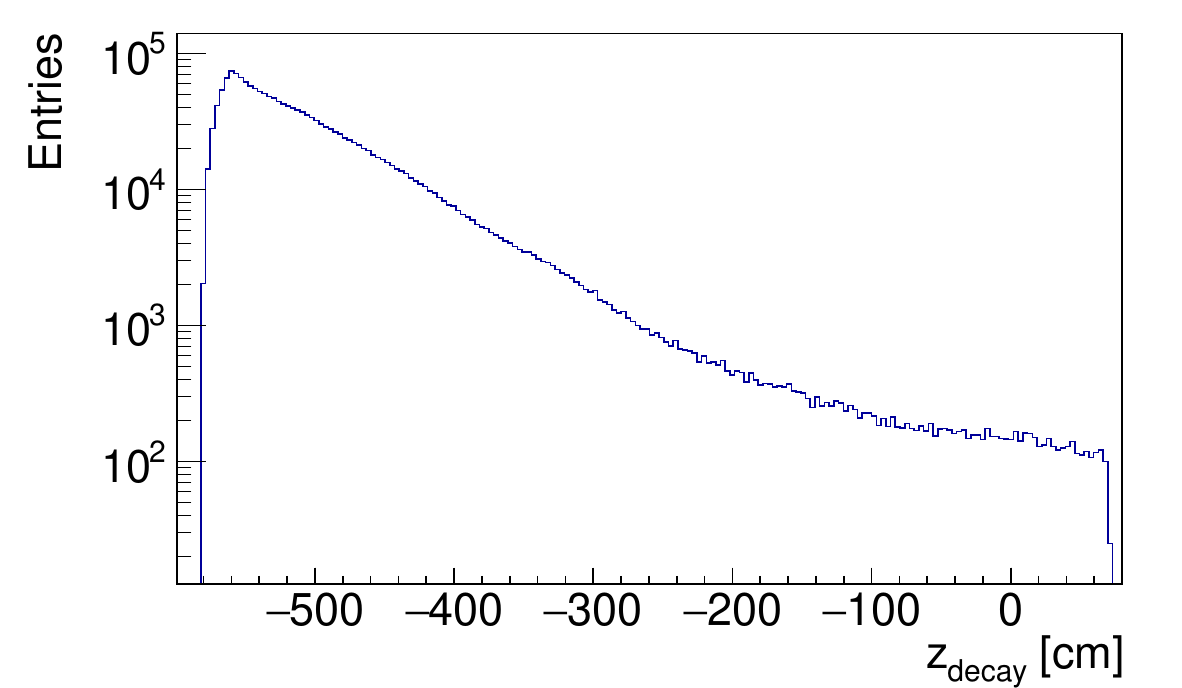}
\end{center}
\caption{
The distribution of the coordinates $z_\text{decay}$ (after cuts). 
}
\label{hist-zdecay}
\end{figure*} 

\begin{figure*}[htbp]
\begin{center}
\includegraphics[width=.6\textwidth]{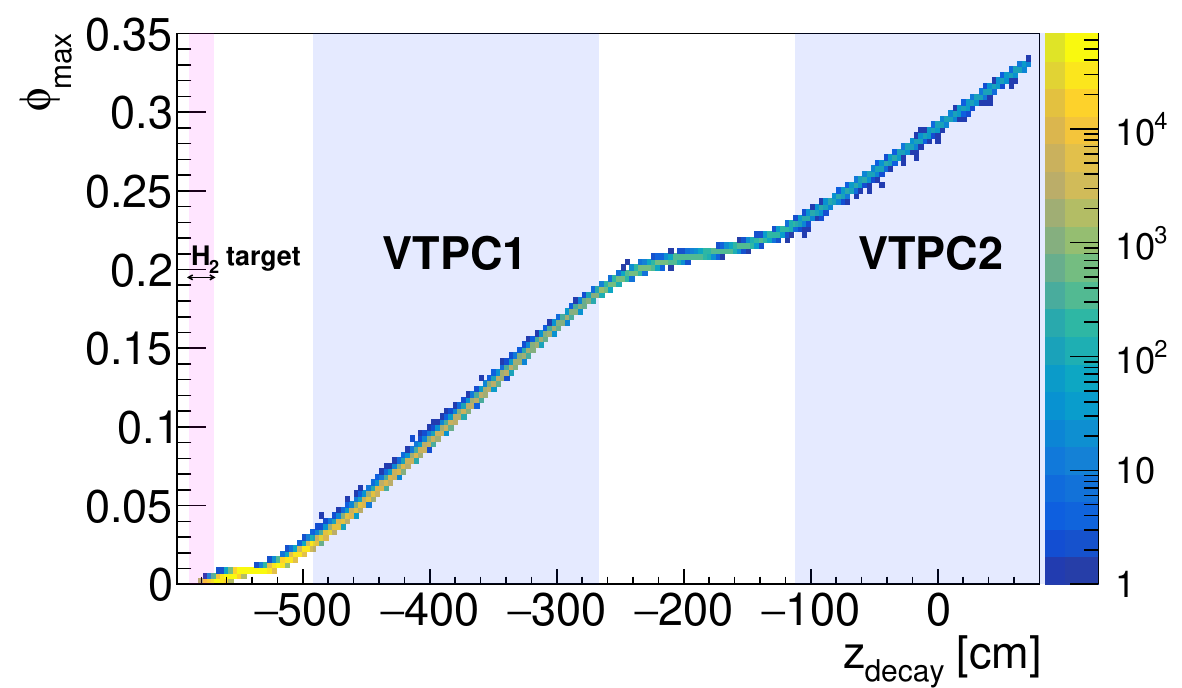}
\end{center}
\caption{Dependence of the maximum precession angle 
$\phi_\text{max}$  on the $z$ coordinate point of $\Lambda$ decay $z_\text{decay}$. 
}
\label{phi_max}
\end{figure*}

The largest value of $z_\text{decay}$ shown in Fig.~\ref{xyzspin-sample} is possible only for most energetic $\Lambda$s, however, taking into account the exponential law of decay, the majority of particles simulated with Epos decays with $z_\text{decay} < -400$~cm, which is shown in Fig.~\ref{hist-zdecay}. Dependence of the corresponding precession angle $\phi_\text{max}$ on position of decay $z_\text{decay}$ is shown in Fig.~\ref{phi_max}. It may reach 0.35, however, for most of the particles, it remains below 0.1.

The same numerical simulations can be used to determine the influence of the magnetic field on the polarization measurement bias. The initial $\Lambda$~polarization vector $\vec{S}$ was generated uniformly in space and propagated in the magnetic field until the particle's decay, according to Eq.~(\ref{spin-precession-dz}). Next, the vector $\vec{S}$  projected along the directions $\hat{n}_x$, $\hat{n}_y$, and $\hat{n}_z$. The distributions of the polarization vector components before and after the evolutions are presented in Figs. 6 and 7. The polarization bias components defined as $P_i=\langle S_i \rangle $  are shown in Fig.~\ref{hist-polarizationbias_xfptbin}. We see an effect smaller than $10^{-2}$, consistent with zero for many of $(x_F,p_T)$ bins.

\begin{figure}[ht]
    \includegraphics[width = 0.325\linewidth]{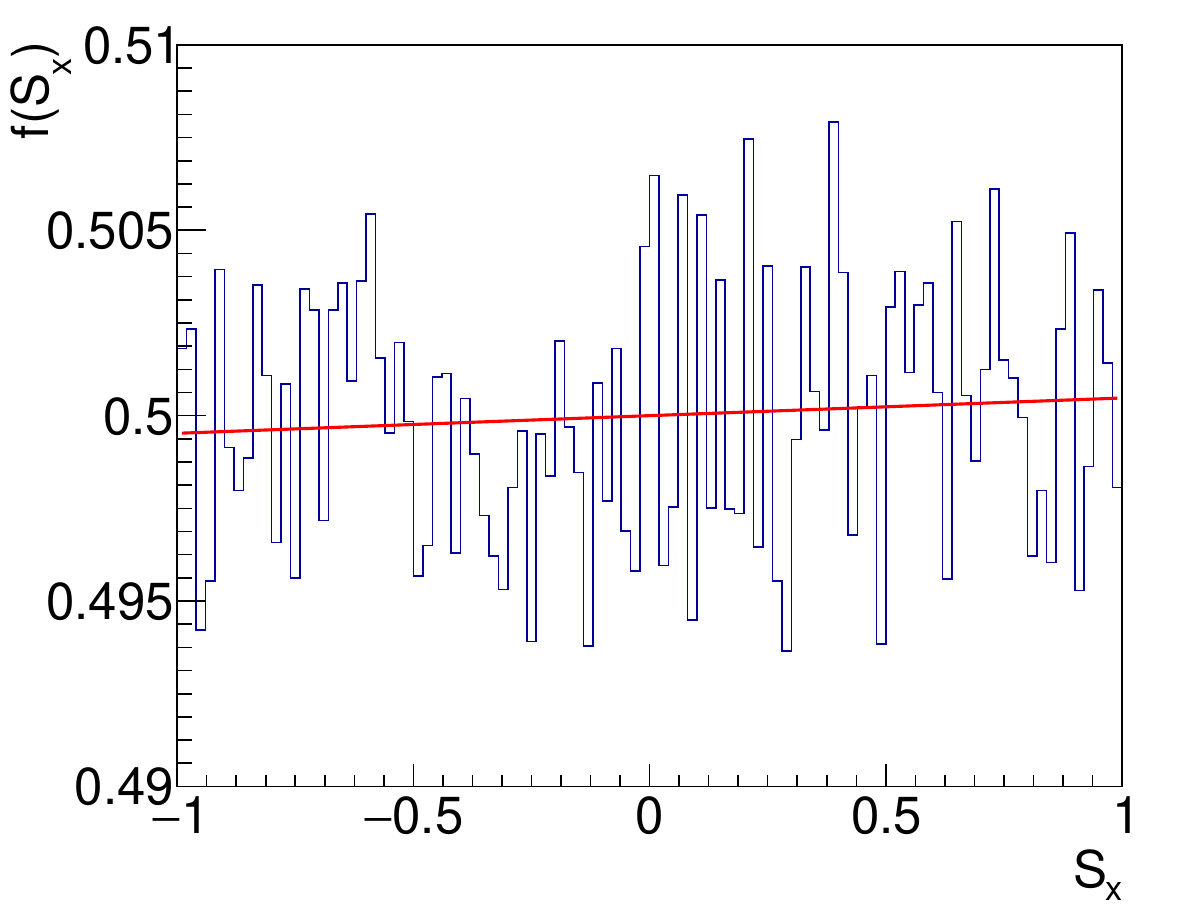}
    \includegraphics[width = 0.325\linewidth]{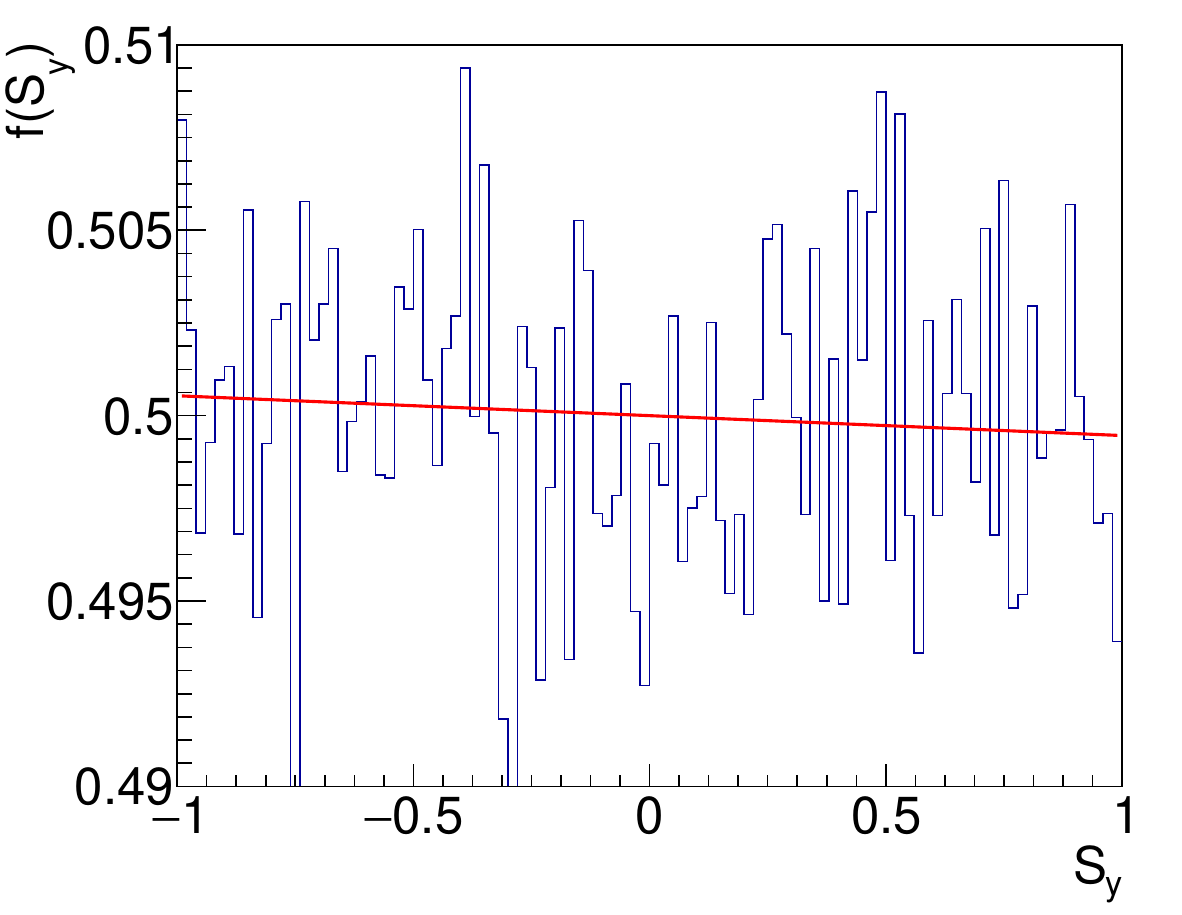}
    \includegraphics[width = 0.325\linewidth]{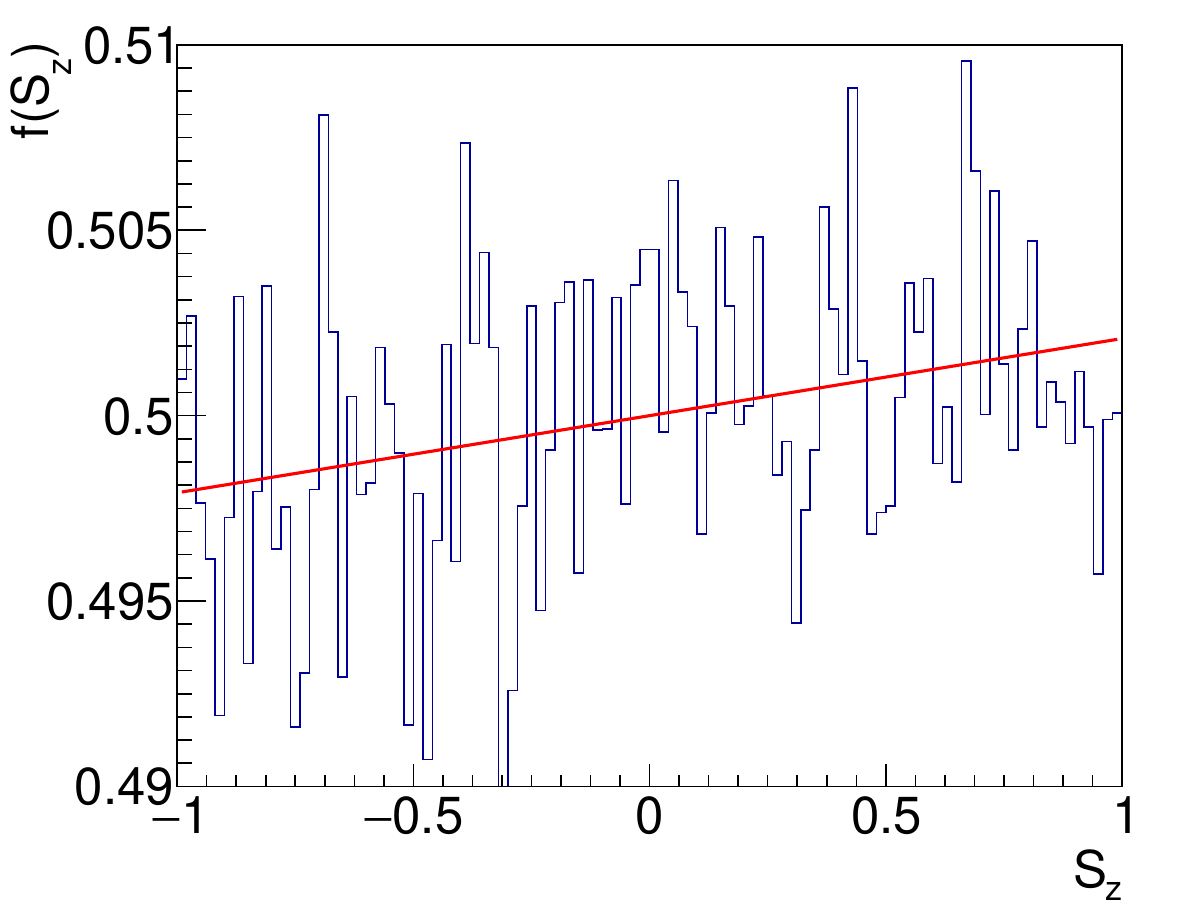}
    \caption{Polarization vector $\vec{S}_\text{init}$ distribution. Fitted values are $P_x = (1.3 \pm 1.8)\cdot10^{-3}$, $P_y = (-1.5 \pm 1.8)\cdot10^{-3}$, and $P_z = (5.7 \pm 1.8)\cdot10^{-3}$.}
    \label{hist-S-init-cut}
    \includegraphics[width = 0.325\linewidth]{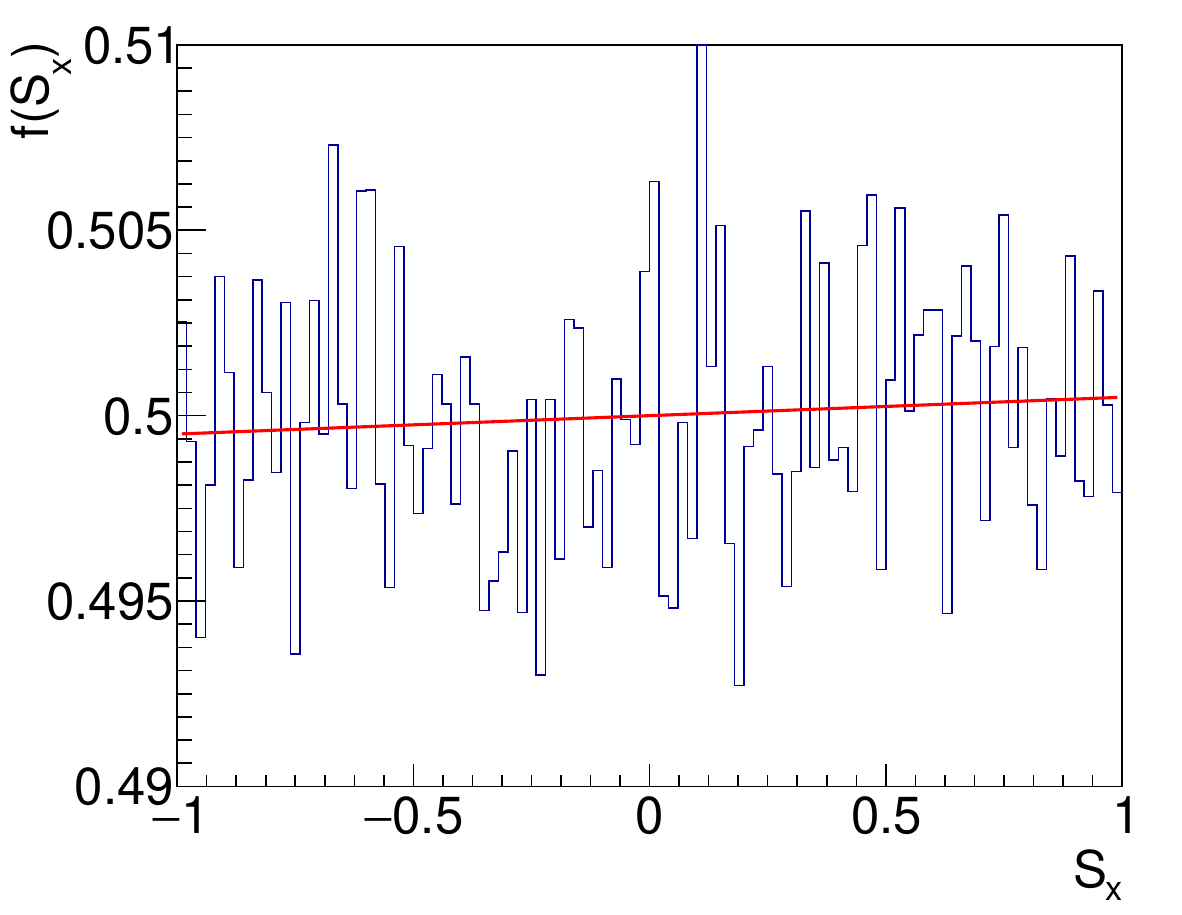}
    \includegraphics[width = 0.325\linewidth]{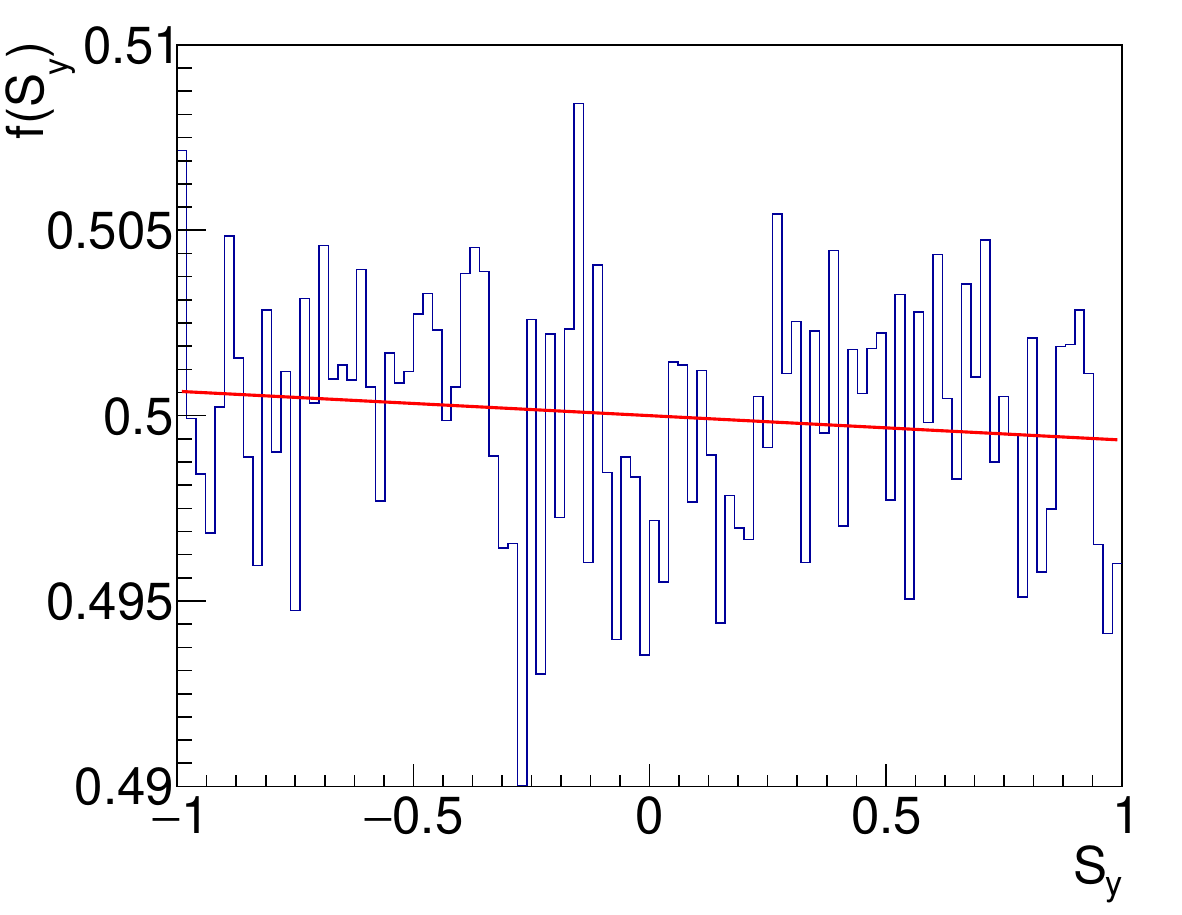}
    \includegraphics[width = 0.325\linewidth]{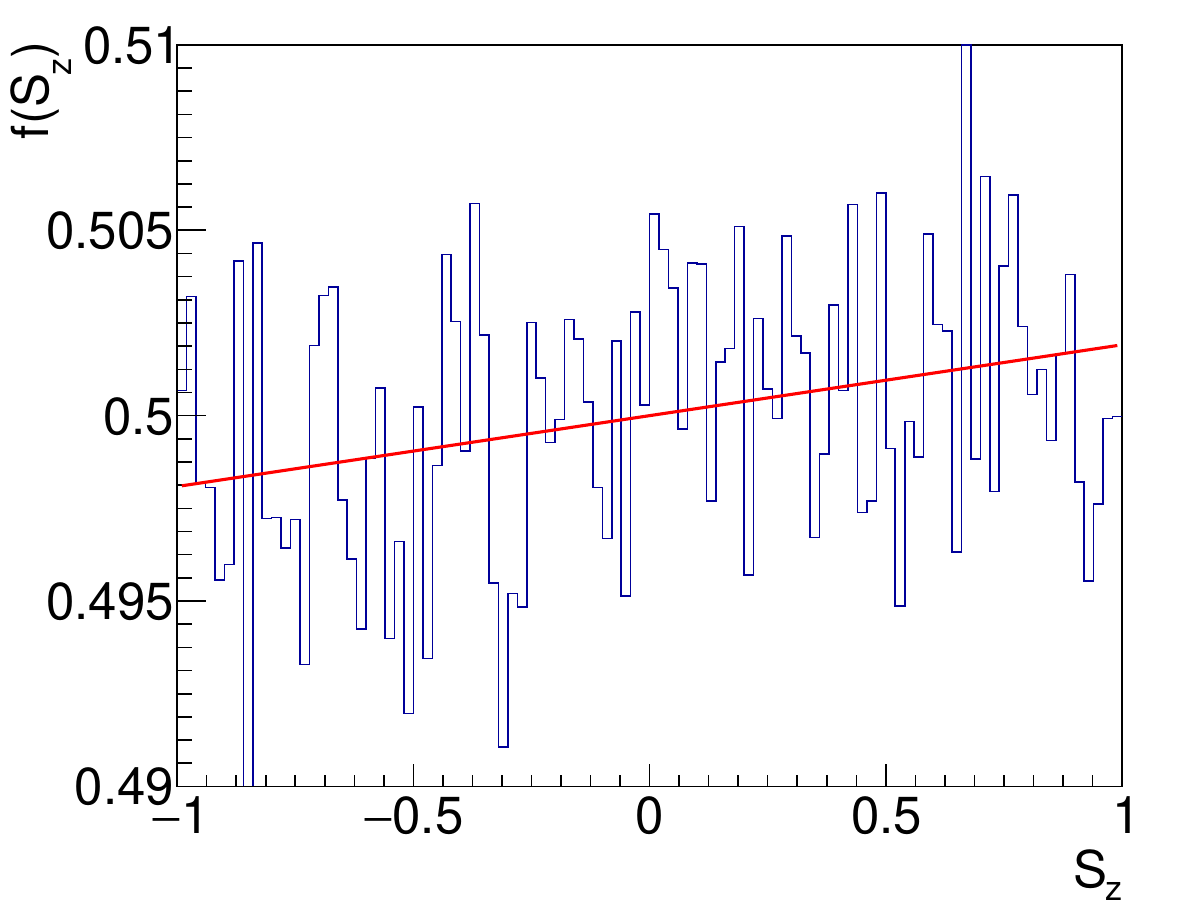}
    \caption{Polarization vector $\vec{S}$ distribution. Fitted values are $P_x = (1.4 \pm 1.8)\cdot10^{-3}$, $P_y = (-1.8 \pm 1.8)\cdot10^{-3}$, and $P_z = (5.2 \pm 1.8)\cdot10^{-3}$.}
    \label{hist-final-cut}
\end{figure}

\section{Summary}

Our main conclusions can be summarized by two points: first, the geometrical acceptance significantly biases the polarization results and should be taken into account via MC corrections, second, the impact of the magnetic field on $\Lambda$ polarization due to precession is at least one order of magnitude smaller than detector acceptance-based polarization bias. 

Consequently, we expect that \NASixtyOne has a large potential to study $\Lambda$ transverse polarization in p--p collisions. The identification of an experimental signal similar to that observed before by other experiments is possible (see, for example, Refs.~\mbox{\cite{PhysRevD.91.032004,ABT2006415,Fantietal.1999,PhysRevD.40.3557,PANAGIOTOU1990}}).

\section*{Acknowledgements}
Y.B. is grateful to the \NASixtyOne collaboration for providing the Monte-Carlo data and computational facilities and M.~Ga\'zdzicki for fruitful discussions. This work was supported in part by the Polish National Science Centre Grant No. 2022/47/B/ST2/01372 (WF) and 2023/49/N/ST2/02299 (YB).

\begin{figure}[t!]
    \includegraphics[width=\linewidth]{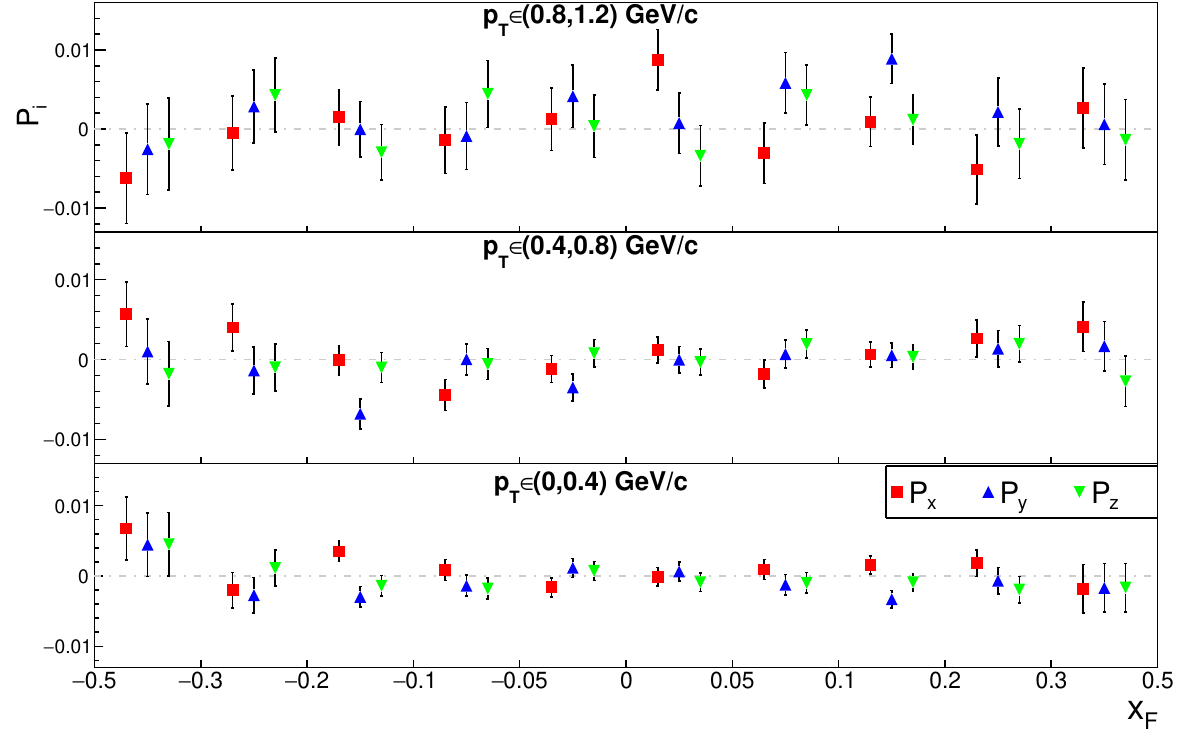}
    \caption{Polarization vector $P_{i}$ distribution for different $(x_F,p_T)$ binning. }
    \label{hist-polarizationbias_xfptbin}
\end{figure}

\section{Appendix: gradient force}

Here we make an estimate of the impact of the force $\vec{F} =  \vec{\nabla} (\vec{m} \cdot \vec{B}) = \sum_k m_k  \vec{\nabla} B_k$ (for $k = x,y,z$, where $\vec{m}$ is particle's magnetic moment) on the polarization evolution. In the laboratory frame, the magnetic field changes by $\Delta B_y \simeq 1.5 \text{ T}$ over the distance $L \simeq 1.5 \text{ m}$ along $z$-axis. Due to time dilation, the particle flight path is $L = \gamma c \tau$, where $\tau $ is the $\Lambda$ mean lifetime and $c\tau = 7.89~\cm$. Hence, $\gamma \simeq 19$ and the hyperon momentum $p_\Lambda \simeq 21~\GeVc$, which is a typical $\Lambda$ momentum for proton-proton collisions at 158 \GeVc beam momentum. 
Numerical value of nuclear magneton is $\mu_N = 3\cdot 10^{-8}\text{ eV/T}$, and the $\Lambda$ magnetic  moment  $\abs{\vec{m}} \approx 0.6 \mu_N$. In the $\Lambda$ rest frame, the momentum change is $\Delta\vec{p} = \vec{F}  \tau =m_y  \vec{\nabla} B_y \tau$. Even if $ \vec{m}$ is aligned with the $y$-axis, we find
\begin{equation}
     \Delta p_z = 0.6\mu_N \frac{\gamma \Delta B_y}{L/\gamma} \tau \approx 5\cdot 10^{-7} \text{ eV}/c.
\end{equation}
Therefore,  $v/c = p_z / (mc) \approx 5\cdot 10^{-16} $ and  this ratio is negligible. From the BMT equation \eqref{jackson-spin-eq}, the spin vector change is quadratic in the speed of particle: $\abs{\Delta \vec{S}} \sim (v/c)^2 \sim 10^{-31}$, hence, the impact of the gradient force can be neglected.

\bibliographystyle{unsrt}
\bibliography{references.bib}
\end{document}